\documentclass[12pt]{article}
\usepackage{graphicx}
\usepackage[cp1251]{inputenc}
\usepackage{epsfig}
\usepackage[english]{babel}

\date{}

\begin{document}
\setcounter{page}{1}
\pagestyle{plain}
\title{\bf{Photo-induced spin and valley-dependent Seebeck effect in the low-buckled Dirac materials}}

\author{Yawar Mohammadi \thanks{Corresponding author's E-mail address:
y.mohammadi@cfu.ac.ir}} \maketitle{\centerline{Department of
physics, Farhangian University, Tehran, Iran}

\begin{abstract}

Employing the Landauer-Buttiker formula we investigate the spin
and valley dependence of Seebeck effect in low-buckled Dirac
materials (LBDMs), whose band structure are modulated by local
application of a gate voltage and off-resonant circularly
polarized light. We calculate the charge, spin and valley Seebeck
coefficients of an irradiated LBDM as functions of electronic
doping, light intensity and the amount of the electric field in
the linear regime. Our calculation reveal that all Seebeck
coefficients always shows an odd features with respect to the
chemical potential. Moreover, we show that, due to the strong
spin-orbit coupling in the LBDMs, the induced thermovoltage in the
irradiated LBDMs is spin polarized, and can also become valley
polarized if the gate voltage is applied too. It is also found
that the valley (spin) polarization of the induced thermovoltage
could be inverted by reversing the circular polarization of light
or reversing the direction the electric field (only by reversing
the circular polarization of light).

\end{abstract}


\vspace{0.5cm}

{\it \emph{Keywords}}: A. Low-buckled Dirac materials; D.
Ballistic transport; D. spin/valley dependent Seebeck effect; D.
off-resonant light irradiation.
%
\section{Introduction}
\label{sec:1}

The low-buckled Dirac materials are monolayer honeycomb lattice
structures of heavy IV-group elements. Such lattice structures
have been synthesized recently for
silicon\cite{Lalmi1,Vogt1,Feng1,Meng1,Fleurence1,Chiappe1},
germanium\cite{Li1,Zhang1}, and also stannum\cite{Zhu1}. They are
also known as silicene, germanene and stanene respectively. In
these structures, atoms arranged in a monolayer honeycomb lattice
which can be described as in graphene in terms of two triangular
sublattices. However, larger ionic size of heavy IV-group atoms
results in buckling of these two-dimensional lattices.
Accordingly, the sites on the two sublattices are shifted
vertically with respect to each other and sit in two parallel
planes with a separation of $0.46~nm$. Due to the puckered
structure, which results in a large spin-orbit
interaction\cite{Liu1,Liu2}, the low energy dynamic in the LBDMs
is dominated by a massive Dirac Hamiltonian\cite{Liu1}, with a
mass which could also be tuned via an electric filed applied
perpendicular to its plane\cite{Drummond1}. These novel features
donate many attractive properties such as quantum spin Hall
effect\cite{Liu1,Xu1}, valley-polarized quantum Hall
effect\cite{Pan1,Ezawa1,Tahir1}, quantum thermal
transport\cite{Zhou1}, topological superconducting
effect\cite{Ezawa2} and so on
\cite{Tabert1,Ezawa3,An1,Tsai1,Ezawa4,Soodchomshom1,Wang2,Missault1,Mohammadi1}
to the LBDMs.

In the LBDMs, the valley and spin degrees of freedom have been
coupled via a spin-orbit interaction\cite{Liu1} which is large
compared with that in graphene\cite{Min1}. Further, they have long
spin-coherence length\cite{Sanvito1} and spin-diffusion
time\cite{Huang1,Wang1}, and also weak inter-valley
coupling\cite{Gunst1}. These properties make them suitable
materials to detect spin- or/and valley-dependent transport
phenomena. Provided that the spin or/and valley degeneracy of
their band structure are lifted\cite{Tabert1,Yokoyama1}. This can
be achieved, for example, in a ferromagnetic LBDM, in which the
magnetic exchange field is supposed to be induced in the LBDM, as
in graphene by an adjacent magnetic insulator such as
$EuO$\cite{Haugen1}, $EuS$\cite{Wei1} or $YIG$\cite{Wang3}. Based
on this fact, recently various spin- and valley dependent
thermoelectric effects, such as spin-valley diode
effect\cite{Zhai1}, spin-valley Seebeck effect\cite{Zhai2} and
anomalous thermospin effect\cite{Gusynin1}, have been predicted
for the ferromagnetic LBDMs. Unfortunately, these effects can be
observed \textit{only for temperature regime lower than magnetic
exchange field(h), namely for $T<h/k_{B}$}. In this letter we
propose a new scheme to achieve the spin- and valley-dependent
Seebeck effects in the LBMDs, which can also overcome the above
mentioned problem of the ferromagnetic LBDMs. Our scheme is based
on a new experimental technique\cite{Wang4,Sie1} which makes it
possible to access effects arising from off-resonant light
irradiation. In the off-resonant regime, light does not directly
excite the electrons, and instead effectively modifies the
electron band structure through virtual photon absorption/emission
processes. Further, it has been shown that the influence of such
off-resonant light is captured in a valley-dependent static
effective Hamiltonian\cite{Ezawa3,Kitgawa1,Tahir2,Mohammadi2}.
This valley-dependent effect, due to the strong spin-valley
coupling in the LBDMs, removes the spin degeneracy of their band
structure and provides a new platform to explore detectable spin-
and valley-dependent transport phenomena in these materials.

In this letter, employing the Landauer–B\"{u}ttiker formalism we
investigate the spin and valley dependence of thermoelectric
effects in an irradiated LBDM-based junction exposed to a
temperature gradient. Our results show that when the central
region of the junction is exposed to off-resonant circularly
polarized light, the energy bands in the central region become
only spin-polarized. This can lead to spin-dependent Seebeck
effect, whose amount and polarization can be controlled by tuning
the light intensity and circulation respectively. If a
perpendicular electric field is applied too, the band structure
becomes spin and valley-polarized. Consequently, the induced
Seebeck effect become spin- and valley-dependent. The valley
(spin) polarization of the induced Seebeck effect can be inverted
by reversing the the circular polarization of light or reversing
the direction of the electric field (only by reversing the the
circular polarization of light). The corresponding Hamiltonian
model and formalism are introduced in sec. II. We present our
numerical results and discussion in sec. III and end the paper by
summary and conclusions in sec. IV.

\section{Model Hamiltonian}
\label{sec:2}

The low energy physics in a low-buckled Dirac material, which is
subjected to a perpendicular electric field, is governed by a
$2\times2$ Hamiltonian matrix as\cite{Liu1,Liu2,Drummond1}
\begin{eqnarray}
H^{\eta,s}=\hbar v_{F}(k_{x}\tau_{x}-\eta k_{y}\tau_{y})-\eta
s\lambda_{so}\tau_{z}+\lambda_{z}\tau_{z}, \label{eq01}
\end{eqnarray}
acting in the sublattice pseudospin space. The first term is the
Dirac Hamiltonian arising from the nearest neighbor transfer
energy in a two-dimensional honeycomb lattice where $\eta=+(-)$
refers to $\mathbf{K}(\mathbf{K^{'}})$ Dirac points and
$v_{F}=\sqrt{3}\gamma a/2\hbar$ is the Fermi velocity with $t$ and
$a$ being the nearest neighbor transfer energy and the lattice
constant which have been listed in table \ref{Tab01} for silicene,
germanene and stanene. Here $\mathbf{k}=(k_{x},k_{y})$ is the two
dimensional momentum measured from Dirac points and
$\tau_{i}(i=x,y,z)$ are Pauli matrixes. The second term is the
Kane-Mele term for the intrinsic spin-orbit coupling, in which
$\lambda_{so}=\lambda_{so1}+\lambda_{so2}$ with $\lambda_{so1}$
and $\lambda_{so2}$ being the first and second order effective
spin-orbit interaction in a low-buckled Dirac material(see table
\ref{Tab01}). In this term $s=+(-)$ index denotes to the spin-up
(-down) degree of freedom. The last term is the on-site potential
difference between A and B sublattices tuned by the perpendicular
electric field, $E_{z}$, in which $\lambda_{z}=\pm edE_{z}$, $e$
is the electron charge, $d$ is the vertical distance between A and
B sublattices, and +(-) is used when $E_{z}$ is along $+z$ ($-z$)
direction.

As mentioned above, the sample is subjected to off-resonant
circularly polarized light, whose vector potential can be written
as $\mathbf{A}(t)=(\pm A \sin\Omega t,A \cos\Omega t)$ where
$\Omega$ is the frequency of light and the plus (minus) sign
denotes to the right (left) circulation. In the off-resonant
regime, satisfied when $\hbar \Omega\gg \gamma$, light does not
directly excite the electrons, and instead effectively modifies
the electron band structure through virtual photon
absorption/emission processes (see Ref. \cite{Kitgawa1}).
Moreover, when the intensity of light is small
($\mathcal{A}=eaA/\hbar\ll 1$), the influence of the off-resonant
light irradiation on the electron band structure (According to the
Flequent theory and the Peierls substitution) is well captured
 by
a static effective
Hamiltonian\cite{Ezawa3,Kitgawa1,Tahir2,Mohammadi2} as $\Delta
H^{\eta,s}=\eta\lambda_{\Omega}\tau_{z}$, where
$\lambda_{\Omega}=\pm (\hbar v_{F}\mathcal{A}/a)^{2}/\hbar \Omega
$ with $+(-)$ for the right (left) circulation of light. Hence,
the low energy physics in low buckled honeycomb structures
subjected to a perpendicular electric field and off-resonant
circularly polarized light is described by an effective
Hamiltonian as
\begin{eqnarray}
H^{\eta,s}_{eff}=\hbar v_{F}(k_{x}\tau_{x}-\eta
k_{y}\tau_{y})-\eta
s\lambda_{so}\tau_{z}+\lambda_{z}\tau_{z}+\eta\lambda_{\Omega}\tau_{z},\label{eq02}
\end{eqnarray}
acting in the sublattice pseudospin space, whose energy bands are
\begin{eqnarray}
\varepsilon^{\eta}_{s}=\nu \sqrt{(\hbar
v_{F}|\mathbf{k}|)^{2}+\Delta^{2}},\label{eq03}
\end{eqnarray}
where
$\Delta=\eta(s_{z}\lambda_{so}-\lambda_{\Omega})-\lambda_{z}$\cite{Ezawa3}
and $\nu=+(-)$ denotes to the conduction (valance) band. The low
energy bands in each region of the junction (the leads and the
conducting region) are obtained from Eq. \ref{eq03}, just by
inserting the corresponding $\lambda_{z}$ and $\lambda_{\Omega}$.

In this letter we study ballistic thermal transport of massive
Dirac fermions through a LBDM-based junction subjected to a
gradient temperature(see Figs. \ref{Fig01} and \ref{Fig05}). The
central region is subjected to off-resonant circularly polarized
light and an electric field applied perpendicular to its plane. We
take x-axis perpendicular to the interfaces and y-axis along them
(see fig. \ref{Fig01}). The interfaces of the leads and the
central region are located at $x=0$ and $x=L$. We also assume the
translational invariance along y-axis satisfied in the limit of
large $W$ ($W$ is the width of the LBDM plane). We restrict our
consideration to $W/L\gg1$ limit, in which the effects of the
microscopic details of the upper and lower edges of the junction
on the electron transport become insignificant\cite{Tworzydlo1}.
According to the generalized Laudauer–B\"{u}ttiker
approach\cite{Bruus1}, the spin and valley-resolved current driven
by temperature difference, $\Delta T=T_{L}-T_{R}$, can be
expressed as
\begin{eqnarray}
I_{\eta,s}=\frac{e}{h}\int_{-\infty}^{-\infty}[n_{F}(E,T_{L})-n_{F}(E,T_{R})]
N(E)T_{\eta,s}(E) dE,\label{eq04}
\end{eqnarray}
in which $N(E)=\frac{W\sqrt{E^{2}-\lambda_{so}^{2}}}{\sqrt{3}\pi t
a}$ is the density of states at the leads,
$n_{F}(E,T_{L(R)})=\frac{1}{[e^{[E-\mu_{L(R)}]/k_{B}T_{L(R)}}+1]}$
is the Fermi-Dirac distribution function of the carriers at
left(right) lead, and $T_{\eta,s}(E)$ can be written in terms of
the valley/spin dependent transmission coefficient as
\begin{eqnarray}
T_{\eta,s}(E)=\int_{-\pi/2}^{+\pi/2}|t_{\eta}^{s}(E,\phi)|^{2}
d\phi,\label{eq05}
\end{eqnarray}
where $\phi=\tan^{-1}(k_{y}/k_{x})$ is the the angle of incidence
and $t_{\eta,s}(E,\phi)$ represents the transmission coefficient
obtained by matching the wave functions and their derivatives at
the interfaces. The wave functions for the valley $\eta$ and the
spin $s$ in the left and right lead can be written respectively as
\begin{eqnarray} \psi^{\eta,s}_{\nu,L}=
e^{ik_{y}y}[\frac{e^{ik_{x}x}}{\sqrt{2\chi_{L}}}\left(
\begin{array}{c}
\sqrt{\chi_{_{L}}-\nu (\eta s_{z}\lambda_{so})}  \\
\nu e^{-i\eta \phi}\sqrt{\chi_{L}+\nu (\eta s_{z}\lambda_{so})}
\end{array}
\right) + r_{\eta,s}\frac{e^{-ik_{x}x}}{\sqrt{2\chi_{L}}}\left(
\begin{array}{c}
\sqrt{\chi_{L}-\nu (\eta s_{z}\lambda_{so})}  \\
\nu e^{-i\eta(\pi-\phi)}\sqrt{\chi_{L}+\nu (\eta
s_{z}\lambda_{so})}
\end{array}
\right)] ,\label{eq06}
\end{eqnarray}
and
\begin{eqnarray} \psi^{\eta,s_{z}}_{\nu,R}=
t_{\eta,s}\frac{e^{i(k_{x}x+k_{y}y})}{\sqrt{2\chi_{L}}}\left(
\begin{array}{c}
\sqrt{\chi_{L}-\nu (\eta s_{z}\lambda_{so})}  \\
\nu e^{-i\eta \phi}\sqrt{\chi_{L}+\nu (\eta s_{z}\lambda_{so})}
\end{array}
\right),\label{eq07}
\end{eqnarray}
respectively, where $\chi_{L}=\sqrt{(\hbar
v_{F}k)^{2}+\lambda_{so}^{2}}$, $r_{\eta,s}$ is the reflection
coefficient and $\nu=+(-)$ denotes to the conduction (valance)
band. The corresponding wave function in the conduction region is
given by
\begin{eqnarray} \psi^{\eta,s}_{\nu,C}=
e^{iq_{y}y}[a_{\eta,s_{z}}e^{iq_{x}x}\left(
\begin{array}{c}
\sqrt{\chi_{C}-\nu \Delta}  \\
\nu e^{-i\eta \theta}\sqrt{\chi_{C}+\nu \Delta}
\end{array}
\right)  +  b_{\eta,s} e^{-iq_{x}x}\left(
\begin{array}{c}
\sqrt{\chi_{C}-\nu \Delta} \\
\nu e^{-i\eta(\pi-\theta)}\sqrt{\chi_{C}+\nu \Delta}
\end{array}
\right)],\label{eq08}
\end{eqnarray}
where $\chi_{C}=\sqrt{(\hbar v_{F}q)^{2}+\Delta^{2}}$ and
$\theta=\tan^{-1}(q_{y}/q_{x})$ is the refractive angle at the
interface. We restrict our calculations to the elastic scattering
at the interfaces, which yields $k\sin \phi=q\sin \theta$, with
$k=(\hbar v_{F})^{-1} \sqrt{E^{2}-\lambda_{so}^{2}}$ and ,
$q=(\hbar v_{F})^{-1} \sqrt{E^{2}-\Delta^{2}}$. The coefficients
$a_{\eta,s_{z}}$ and $b_{\eta,s_{z}}$ could be determined by
requiring continuity of the wave functions. Matching the wave
functions at the interfaces yields
\begin{eqnarray}
t_{\eta,s_{z}}=(\cos(q_{x}L)-i
\mathcal{F}(E,\phi)\sin(q_{x}L))^{-1}, \label{eq09}
\end{eqnarray}
for the transmission coefficient in which
\begin{eqnarray}
\mathcal{F}(E,\phi)=\frac{k_{x}
^{2}\varepsilon_{C}^{2}+q_{x}^{2}\varepsilon_{L}^{2}
+k_{y}^{2}(\varepsilon_{C}-\varepsilon_{L})^{2}}{2\varepsilon_{L}\varepsilon_{C}k_{x}q_{x}},
\label{eq10}
\end{eqnarray}
where $\varepsilon_{L}=|E|+\lambda_{so}$ and
$\varepsilon_{C}=|E|+|\Delta|$.

Employing the linear response assumption\cite{Alomar1}, i.e.,
$T_{L} \approx T_{R}$ = T, we obtain from Eq. \ref{eq04} the spin-
and valley-resolved thermopower as
\begin{eqnarray}
S_{\eta,s_{z}}=-\frac{1}{eT}\frac{L^{1}_{\eta,s}}{L^{0}_{\eta,s}},
\label{eq11}
\end{eqnarray}
with $L^{n(=0,1)}_{\eta,s}=\frac{1}{\hbar}\int dE (E-\mu)^{n}\int
d\alpha \cos\alpha T_{\eta,s}(E)[-\partial_{E}f(E)]$. In analogy
with the charge thermopower,
$S_{c}=\sum_{\eta,s}S_{\eta,s_{z}}/2$, one can define the spin-
and valley-dependent thermopower, calculated as \cite{Niu1}
$S_{s}=\sum_{\eta,s}s S_{\eta,s_{z}}$ and $S_{s}=\sum_{\eta,s}\eta
S_{\eta,s_{z}}$ respectively. The spin- and/or valley-dependent
Seebeck effect is observed when $S_{s}$ and/or $S_{v}$ become
nonzero. The corresponding numerical results are presented in the
next section.

\section{Results and Discussion}
\label{sec:3}

In this section we present our numerical results for the
photo-induced spin- and valley-dependent Seebeck effect in the
LBDM-based junctions. This will be done by calculating the charge,
spin and valley Seebeck coefficients ($S_{c}$, $S_{s}$ and
$S_{v}$) for an irradiated LBDM-based junction. We present our
results in two cases. In the first one, the central region of the
LBDM-junction is subjected only to off-resonant circularly
polarized light, but in the second one it is simultaneously
exposed to off-resonant circularly polarized light and a vertical
electric field.

First we explain the general features of the Seebeck effect in the
LBDM-based junctions. The LBDMs are semiconductors with small band
gaps. So temperature can play the role of exciting
thermo-electrons in the LBDMs, and both electrons and holes can
contribute to their transport phenomena even in the undoped
regime. Additionally, according to Eq. \ref{eq11}, to induce
Seebeck effect in a LBDM, the number of the transmission modes
must be asymmetric about the Fermi energy. In our scheme, this is
satisfied only when LBDM becomes electron- or hole-doped, which
can be done by a back gate voltage\cite{Novoselov01} or by
depositing alkali metals on the surface\cite{Praveen1} as in
graphene. In the doped regime the thermally activated electrons
(hole), which move in the direction (the opposite direction) of
the temperature gradient (see Eq. \ref{eq04}), result in different
charge accumulations at both sides of the junction, namely
$S_{c}\neq 0$. So, depending on whether the LBDM-based junction is
electron or hole doped, the charge thermopower becomes positive or
negative. This can be seen in the panel (a) of Fig. \ref{Fig02}
(the black curve), in which we exhibit the charge thermopower of a
LBDM-based junction as a function of \textit{scaled chemical
potential}, $\mu/\lambda_{so}$. Furthermore, the charge
thermopower shows an odd feature respect to the chemical
potential, namely $S_{c}(-\mu)=-S_{c}(\mu)$. This is similar to
the well-known effect in semiconductors in which the thermopower
for n and p-type ones has opposite sign. However, in a LBDM-based
junction the sign of the thermopower can be changed easily by
tuning the chemical potential. Moreover, one can see that, in the
absence of off-resonant light irradiation ($\lambda_{\Omega}=0$),
the induced Seebeck effect is not spin polarized (see the black
curve of Fig. \ref{Fig03}(b)). This is due to the spin degeneracy
of the band structure of the LBDMs, which leads to same
accumulation for both spin-up and spin-down charge carriers at
both sides of the junction. Note, although the spin degeneracy in
each valley can be lifted via a vertical electric filed, this can
not induce spin-dependent transport effect\cite{Tsai1}.

In our proposed method, to achieve spin-dependent Seebeck effect,
the central region of the LBDM-based junction is irradiated by
off-resonant circularly polarized light (see Fig. \ref{Fig01}).
Consequently, as discussed above, light modifies the energy bands
through virtual photon absorption/emission processes, making them
spin polarized with same polarization at $\mathbf{K}$ and
$\mathbf{K}^{'}$ valley (see panel (c) and (d) of Fig.
\ref{Fig01}). This can lead to a spin-dependent Seebeck effect as
shown in the panel (b) of Fig. \ref{Fig01}, in which the spin
thermopower of a LBDM-based junction have been drawn as a function
of the scaled chemical potential for different $\lambda_{\Omega}$
at temperature $T=0.4\lambda_{so}/k_{B}$. It is evident that the
spin thermopower, like the charge thermopower, shows an odd
feature respect to the chemical potential. However, the charge and
the spin thermopower have opposite sign at same chemical
potential. Moreover, one can see that both the spin and charge
thermopowers can be enhanced by increasing the light intensity
(increasing $\lambda_{\Omega}$).

Figure \ref{Fig03} shows the charge and spin thermopower of an
irradiated LBDM-based junction as a function of the scaled
chemical potential for different temperatures. It is evident that
the amounts of $S_{c}$ and $S_{s}$ decrease by increase of
temperature. This can be explained as follows; At low temperatures
the energy broadening of the derivative of Fermi-Dirac
distribution function, that determines which electrons or holes
could contribute to the Seebeck effect, is small. So, for  $\mu>0$
($\mu<0$) the Seebeck effect is mostly induced by thermally
activated electrons (holes) with special spin index. By increasing
the temperature of the junction, the the energy broadening of
$\partial_{E}f(E)$ increases and it overlaps with the other energy
bands, which results in $S_{c}$ and/or $S_{s}$ with opposite sign.
Hence, the induced spin and charge thermovoltage decreases.
Moreover, one can see that even at hight temperatures, at which
the exchange interaction induced in the ferromagnetic LBDM-based
junctions will be quenched, the amount of the spin thermopower of
the irradiated LBDM-based junction is large. Based on numerical
results, it is about $0.15 mV/K$ at 36K, 430K and 600K for
silicene, germanene and stanene respectively, which could be
detected easily (see Fig. \ref{Fig04} (b)). \textit{This is an
important advantage of irradiated LBDM-based junctions over the
ferromagnetic LBDM-based ones}.

Another important advantage of irradiated LBDM-based junctions
over ferromagnetic LBDM-based ones is that \textit{the induced
spin-polarized thermovoltage in the LBDM-based junctions could be
inverted easily by reversing the circular polarization of light
(determined by the sign of $\lambda_{\Omega}$)}. This can be
understood based on the effect of the off-resonant light
irradiation on their band structure (see the (c) and (d) panels of
Fig. \ref{Fig01}); Changing the circular polarization of light,
from right to left polarization or vise versa, interchanges the
spin-up and spin-down electron bands in the central region.
Consequently, the spin polarization of the induced thermovoltage
becomes inverted. This can be seen in the Fig. \ref{Fig04}, in
which we exhibit the charge and spin thermopower of a LBDM-based
junction as a function of $\lambda_{\Omega}$. Moreover, this
figure show that the charge thermopower is an even function of the
$\lambda_{\Omega}$. This is due to this fact that changing the
circulation of light just interchanges the spin-up and spin-down
energy bands.

To achieve both the spin- and valley-dependent Seebeck effects at
the same time, the central region of the LBDM-based junction must
be simultaneously subjected to off-resonant circularly polarized
light and a vertical electric filed. Hence, the band structure of
the central region is modified through virtual photon
absorption/emission processes and also by the vertical electric
field. This leads to a new energy band which, due to the strong
spin-valley coupling in the LBDMs, is spin- and valley polarized.
See Fig. \ref{Fig05}. This can lead to spin- and valley-dependent
Seebeck effects. The valley (spin) polarization of the induced
thermovoltage could be inverted by reversing the circular
polarization of light or the direction the electric field (only by
reversing the circular polarization of light). Because changing
the circular polarization of light or the direction of the
electric field (only changing the circular polarization of light),
invert the valley (spin) polarization of the energy bands in the
central region of the junction (Fig. \ref{Fig05} (b)-(e)).

Figures \ref{Fig06} and \ref{Fig07} show the charge and spin
thermopower of the irradiated LBDM-based junction exposed to the
vertical electric filed as functions of
$\lambda_{\Omega}/\lambda_{so}$ and $\lambda_{z}/\lambda_{so}$. In
the results presented in the panel (a) and (b) the chemical
potential is $-0.5\lambda_{so}$ and $+0.5\lambda_{so}$
respectively. These figures, in addition to confirming our
previous results \textit{i.e.} $S_{c(s)}(-\mu)=-S_{c(s)}(\mu)$,
$S_{c}(-\lambda_{\Omega})=S_{c}(\lambda_{\Omega})$ and
$S_{s}(-\lambda_{\Omega})=-S_{s}(\lambda_{\Omega})$, show that
$S_{c}$ and $S_{s}$ are even in $\lambda_{z}$, namely reversing
the direction of the electric filed doesn't change the spin
polarization of the induced thermovoltage. This can be understood
based on the effect of the vertical electric field on the band
structure of the junction (see Fig \ref{Fig05} (b)-(e)).

In Fig. \ref{Fig08} we have exhibited the same plots as in Figs.
\ref{Fig06} and \ref{Fig07} but for the valley thermopwer. Figure
\ref{Fig08} shows that: Firstly, when both the electric field and
the light irradiation are present ($\lambda_{\Omega}\neq 0$ and
$\lambda_{z}\neq 0$) the induced Seebeck effect become valley
dependent too, namely $S_{v}\neq 0$. Secondly, by only applying a
vertical electric field the Seebeck effect doesn't occur. Thirdly,
like $S_{c}$ and $S_{s}$, the valley thermopower shows an odd
feature respect to the chemical potential. \textit{Fourthly, one
can invert the valley polarization of the induced thermovoltage by
reversing the circular polarization of light or reversing the
direction of the electric field,
$S_{v}(-\lambda_{\Omega})=-S_{v}(\lambda_{\Omega})$ and
$S_{v}(-\lambda_{z})=-S_{v}(\lambda_{z})$. }

At the end a point is worth of mentioning. Our calculations (don't
presented here) reveal that the plots of Seebeck coefficients of
silicene, germanene and stanene, except for a minor difference at
the amount of them for silicene, are equal and show same features.
So, we presented our numerical results, in general, for a
LBDM-based junction, and we didn't exhibit the plots for each of
them separately.

\section{Summary and conclusions}
\label{sec:4}

In summary we studied the spin and valley dependence of Seebeck
effect in a LBDM-based junction, whose energy bands was assumed to
be modulated by local application of a gate voltage and
off-resonant circularly polarized light. Employing the
Landauer-Buttiker formula we calculated the charge, spin and
valley Seebeck coefficients in the linear regime. Our calculation
revealed that when the junction is only subjected to the light
beam, the induced Seebeck effect is spin polarized. We showed that
the spin polarization of the induced thermovoltage could be
controlled by tuning the light intensity and its circulation.
Moreover we found if off-resonant circularly polarized light and
the vertical electric field are applied simultaneously, the
induced Seebeck effect becomes valley polarized too. It was also
shown that the valley (spin) polarization of the induced
thermovoltage could be inverted by reversing the circular
polarization of light or the direction the electric field (only by
reversing the circular polarization of light). The proposed
thermoelectric device has two important advantages over the
ferromagnetic one. The first is that it can operate at high
temperatures at which the exchange interaction induced in the
ferromagnetic LBDM-based junctions will be quenched. Another
important advantage is that the induced spin and valley polarized
thermovoltages in the irradiated LBDM-based junctions could be
inverted easily by reversing the circular polarization of light or
reversing the direction of the applied electric field.

 \nonumber \section{acknowledgment} This work was
supported by farhangian university.

%

%
%
%
\newpage
\begin{table}
\caption{The table presents geometrical and electrical parameters
($a$, $\lambda_{so1}$, $\lambda_{so2}$, $v_{F}$ and $\gamma$) of
silicene (Si), germanene (Ge) and stanene (Sn) taken from Ref.
\cite{Liu2}. $\gamma$ has been obtained using Eq. 43 of Ref.
\cite{Liu2}. }\label{Tab01}
  \centering
  \begin{tabular}  { c  c  c  c  c c }\hline \hline
 system  & $a$(\AA) & $\lambda_{so1}(meV)$
   & $\lambda_{so1}(meV)$ & $v_{F}$($10^{5}m/s$)  & $\gamma$(eV) \\ \hline \hline
   Silicene & $3.86$ & $3.9$ & $0.073$ & $5.52$  & 1.043 \\\hline
  Germanene  & $4.02$ & $43$ & $3.3$ & $4.57$  & 0.864 \\\hline
 Stanene & $4.70$ & $29.9$ & $34.9$ & $4.85$ & 0.784 \\\hline
\end{tabular}
\end{table}
\begin{figure}
\begin{center}
\includegraphics[width=12cm,angle=0]{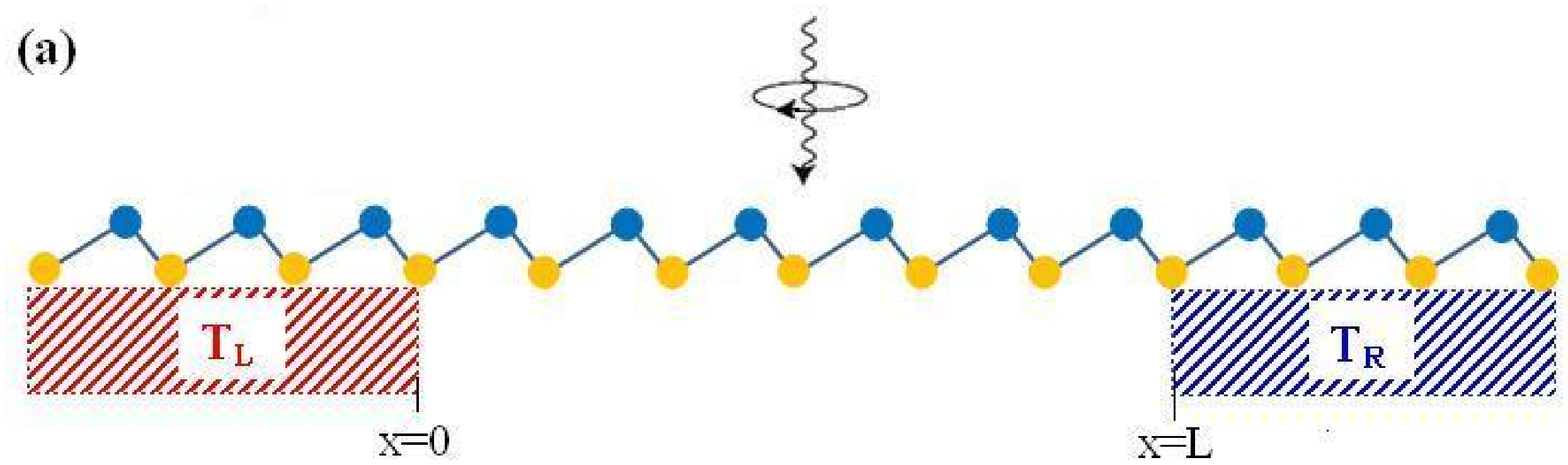}
\includegraphics[width=12cm,angle=0]{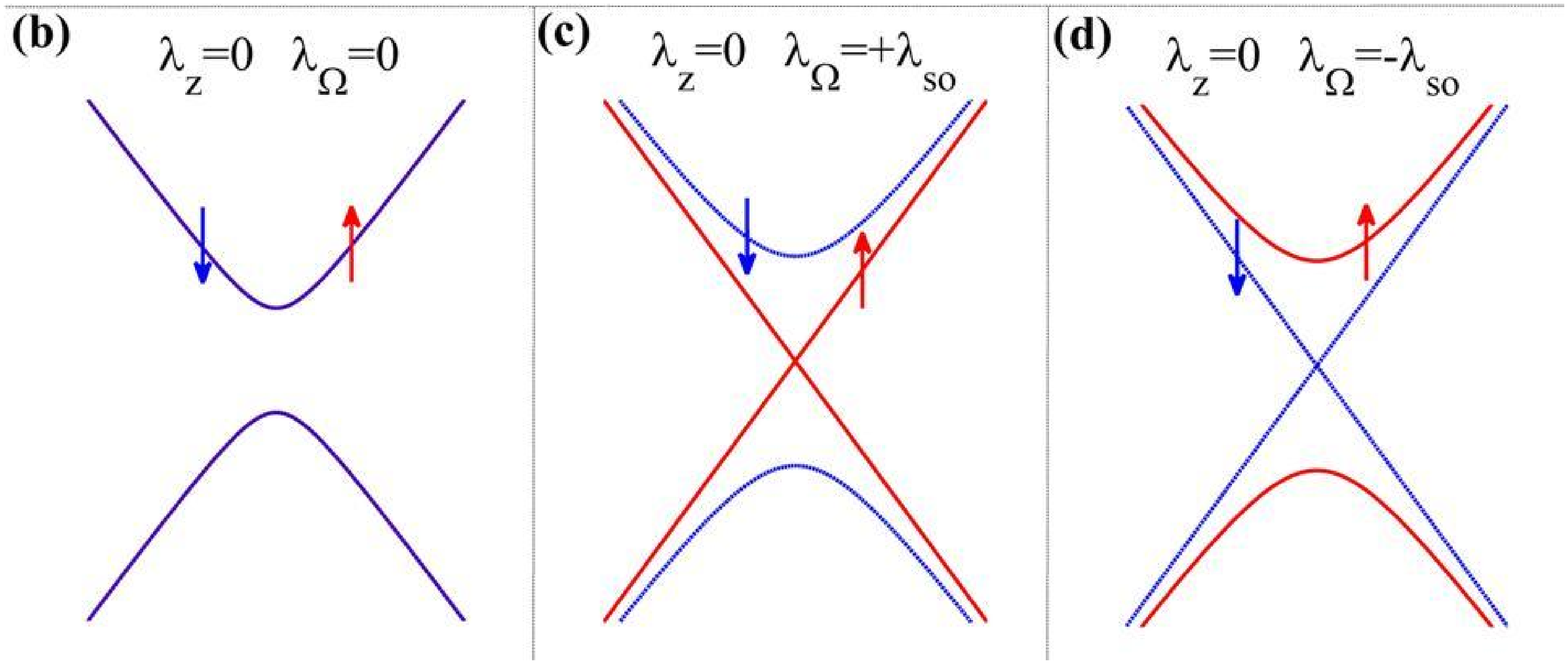}
\caption{(a) A schematic diagram of a LBDM-based junction, whose
central region is subjected to off-resonant circularly polarized
light. Electronic transport is activated with a temperature
gradient ($T_{L}-T_{R}$) between the two hot($L$) and cold($R$)
electrodes. (b), (c) and (d) represent the energy spectrum of the
spin-up (red curve) and spin-down (blue curve) energy bands in the
central region for $\lambda_{\Omega}=0$,
$\lambda_{\Omega}=+\lambda_{so}$ and
$\lambda_{\Omega}=-\lambda_{so}$ respectively}\label{Fig01}
\end{center}
\end{figure}
\begin{figure}
\begin{center}
\includegraphics[width=15cm,angle=0]{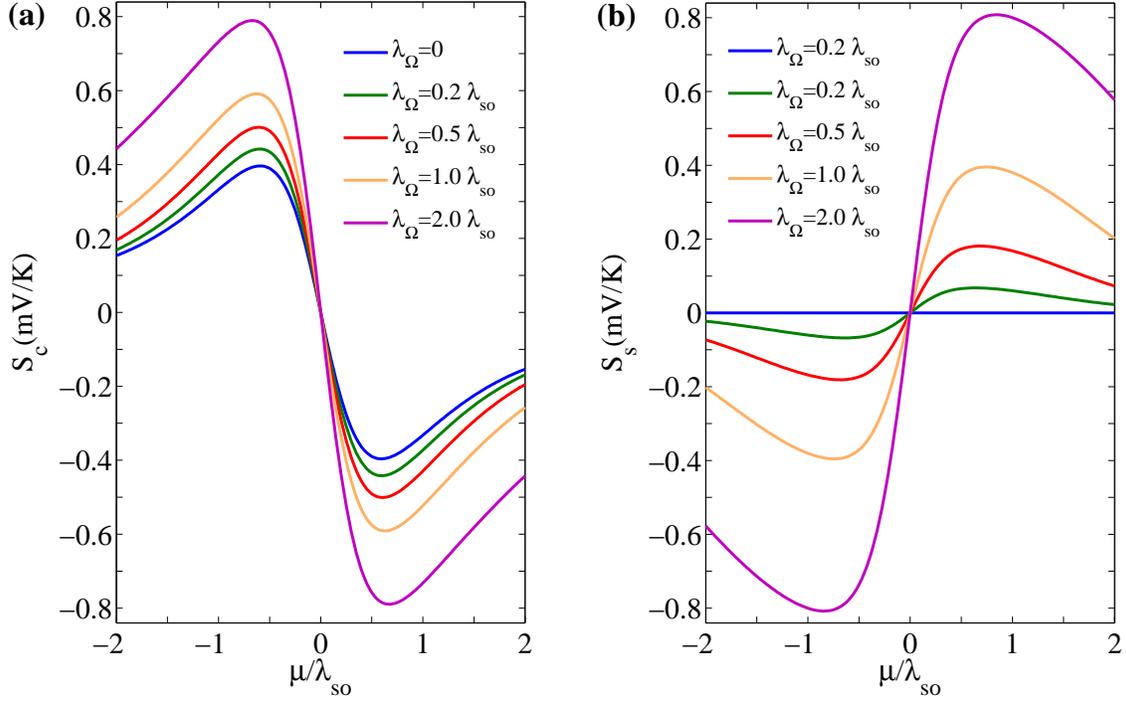}
\caption{(a) Charge and (b) spin thermopower of an irradiated
LBDM-based junction as functions of dimensionless chemical
potential for different values of $\lambda_{\Omega}$. The other
parameters are $\lambda_{z}=0$, $T=0.4\lambda_{so}$ and $L=130
a$.}\label{Fig02}
\end{center}
\end{figure}
\begin{figure}
\begin{center}
\includegraphics[width=15cm,angle=0]{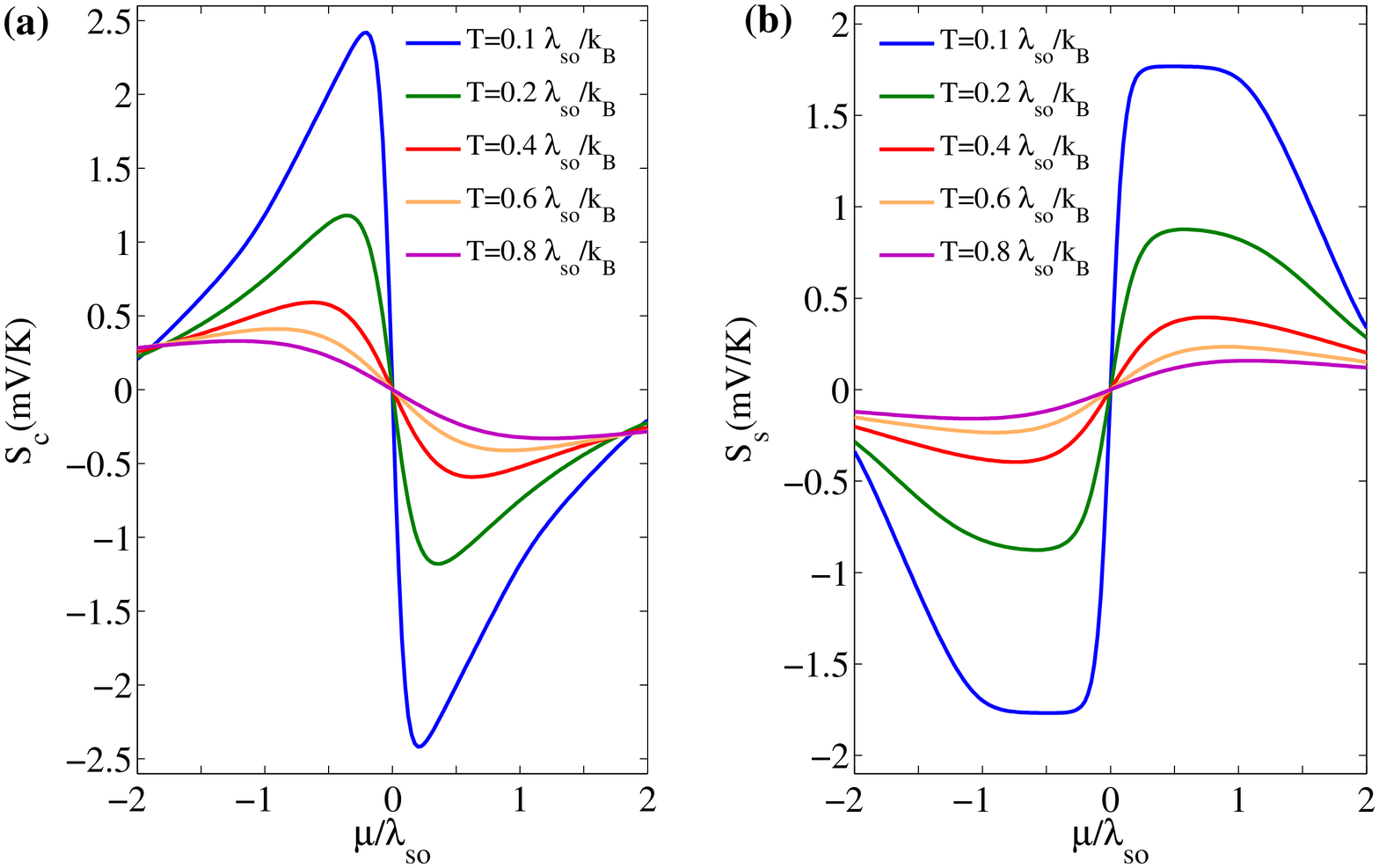}
\caption{(a) Charge and (b) spin thermopower of an irradiated
LBDM-based junction as functions of dimensionless chemical
potential for different values the junction temperature. The other
parameters are $\lambda_{z}=0$, $\lambda_{\Omega}=\lambda_{so}$
and $L=130 a$.}\label{Fig03}
\end{center}
\end{figure}
\begin{figure}
\begin{center}
\includegraphics[width=15cm,angle=0]{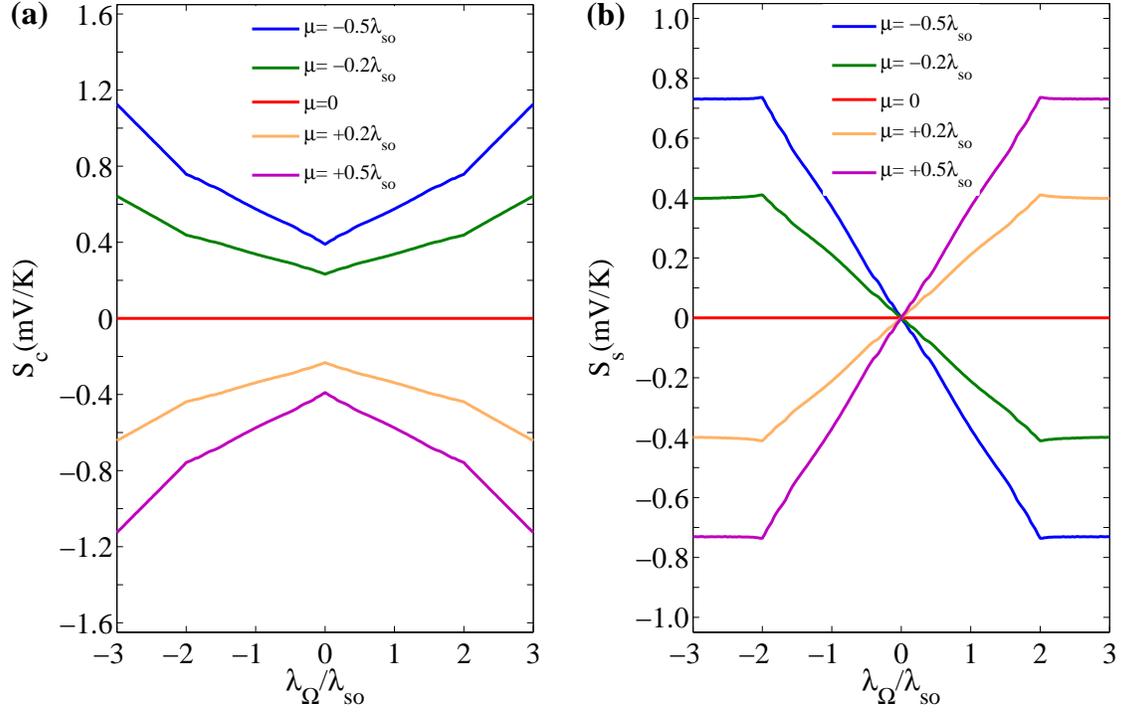}
\caption{(a) Charge and (b) spin thermopower of an irradiated
LBDM-based junction as functions of
$\lambda_{\Omega}/\lambda_{so}$ for different values of the
chemical potential. The other parameters are $\lambda_{z}=0$,
$T=0.4\lambda_{so}$ and $L=130 a$.}\label{Fig04}
\end{center}
\end{figure}
\begin{figure}
\begin{center}
\includegraphics[width=12cm,angle=0]{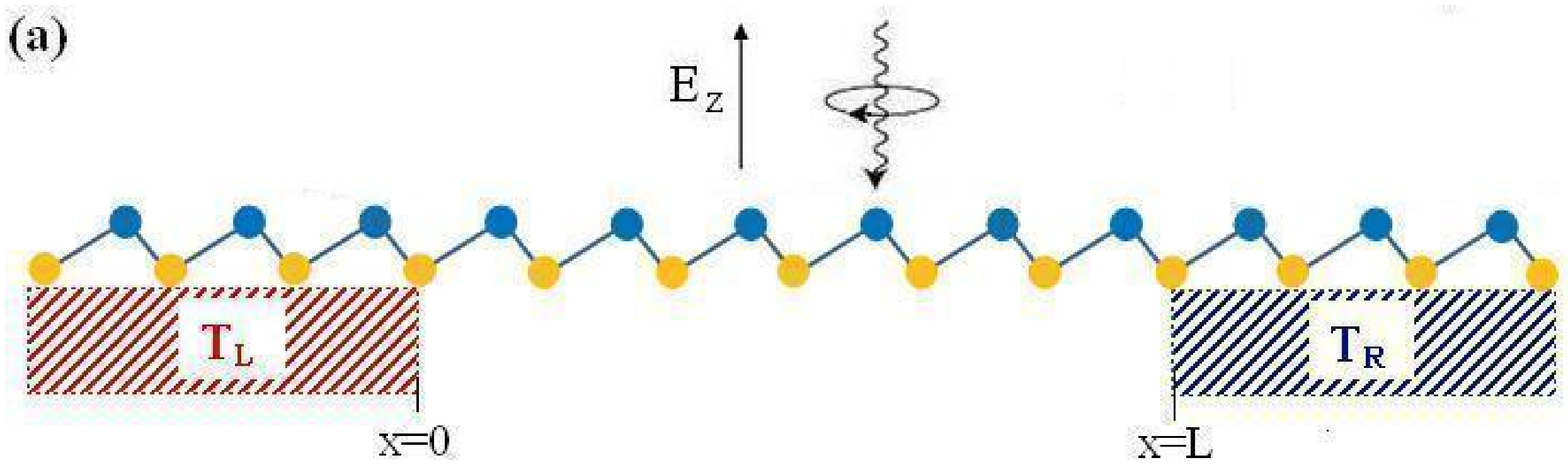}
\includegraphics[width=12cm,angle=0]{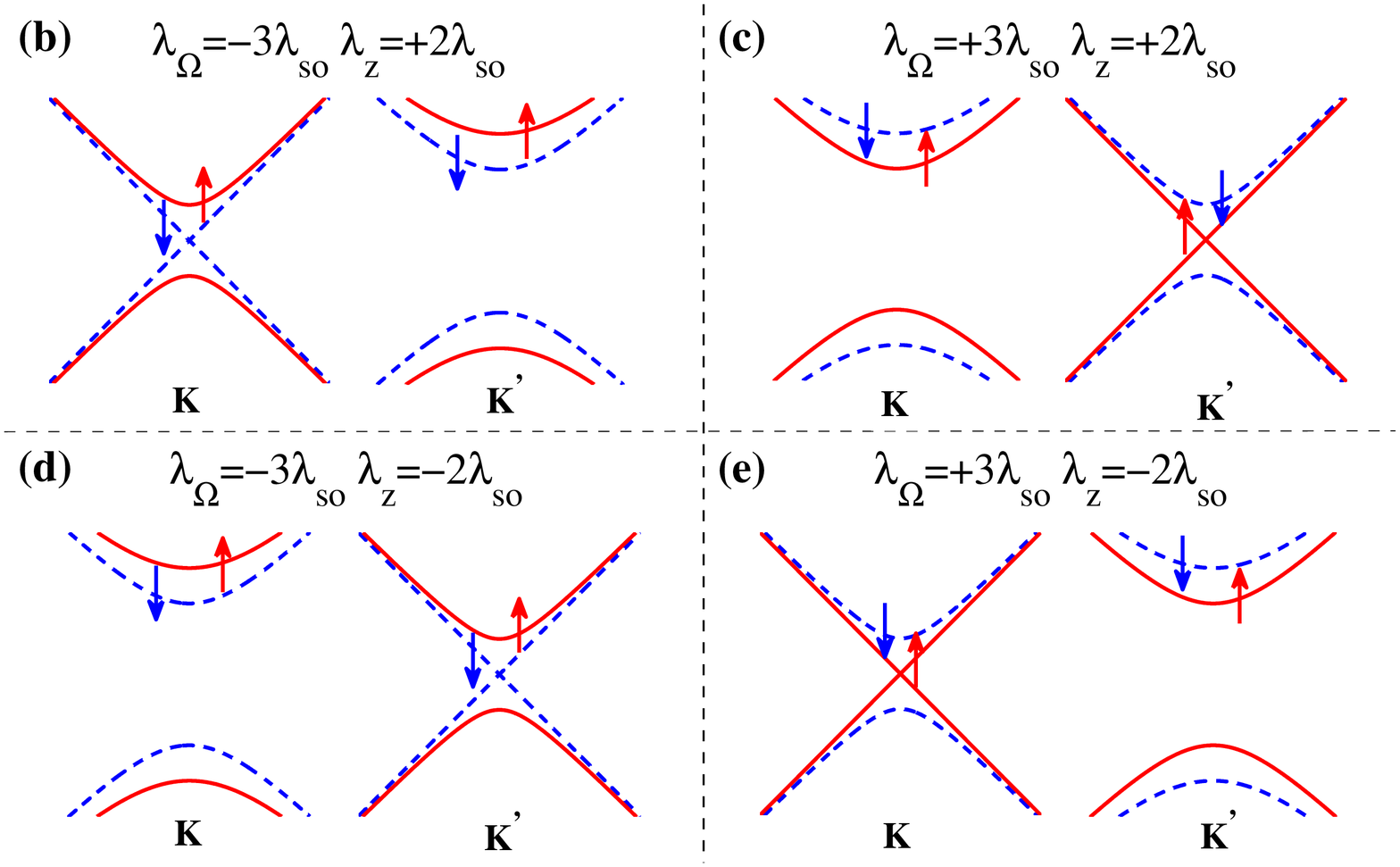}
\caption{(a) A schematic diagram of a LBDM-based junction, whose
central region is subjected to a perpendicular electric field and
off-resonant circularly polarized light simultaneously. (b), (c),
(d) and (e) represent the energy spectrum of the central region
for the spin-up (red curve) and spin-down (blue curve) electrons
at both Dirac points for different values of $\lambda_{\Omega}$
and $\lambda_{z}$.}\label{Fig05}
\end{center}
\end{figure}
\begin{figure}
\begin{center}
\includegraphics[width=15cm,angle=0]{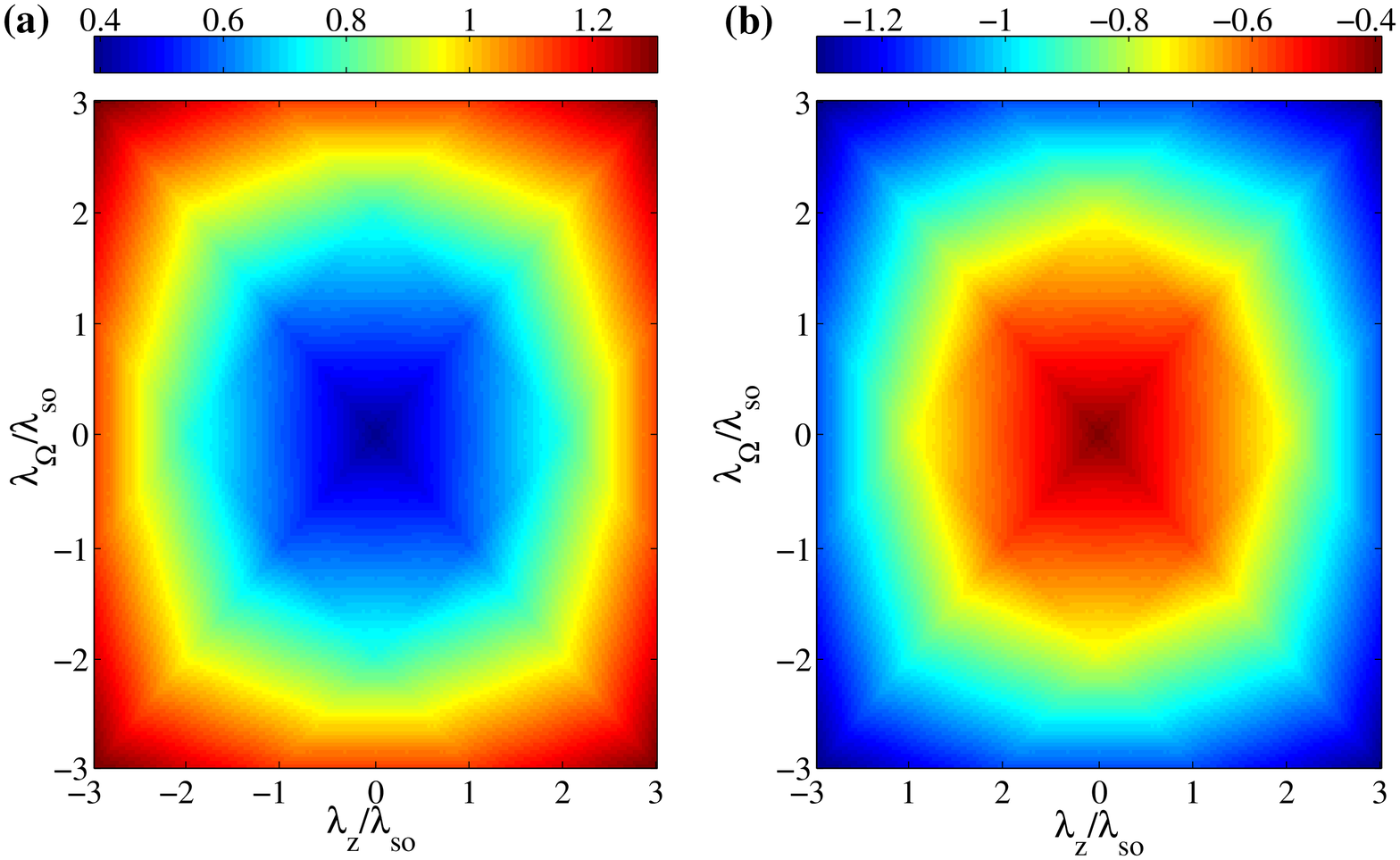}
\caption{Charge thermopower of an irradiated LBDM-based junction
as functions of $\lambda_{\Omega}/\lambda_{so}$ and
$\lambda_{z}/\lambda_{so}$ for (a) $\mu=-0.5\lambda_{so}$ and (b)
$\mu=+0.5\lambda_{so}$. The other parameters are
$T=0.4\lambda_{so}/k_{B}$ and $L=130a$.}\label{Fig06}
\end{center}
\end{figure}
\begin{figure}
\begin{center}
\includegraphics[width=15cm,angle=0]{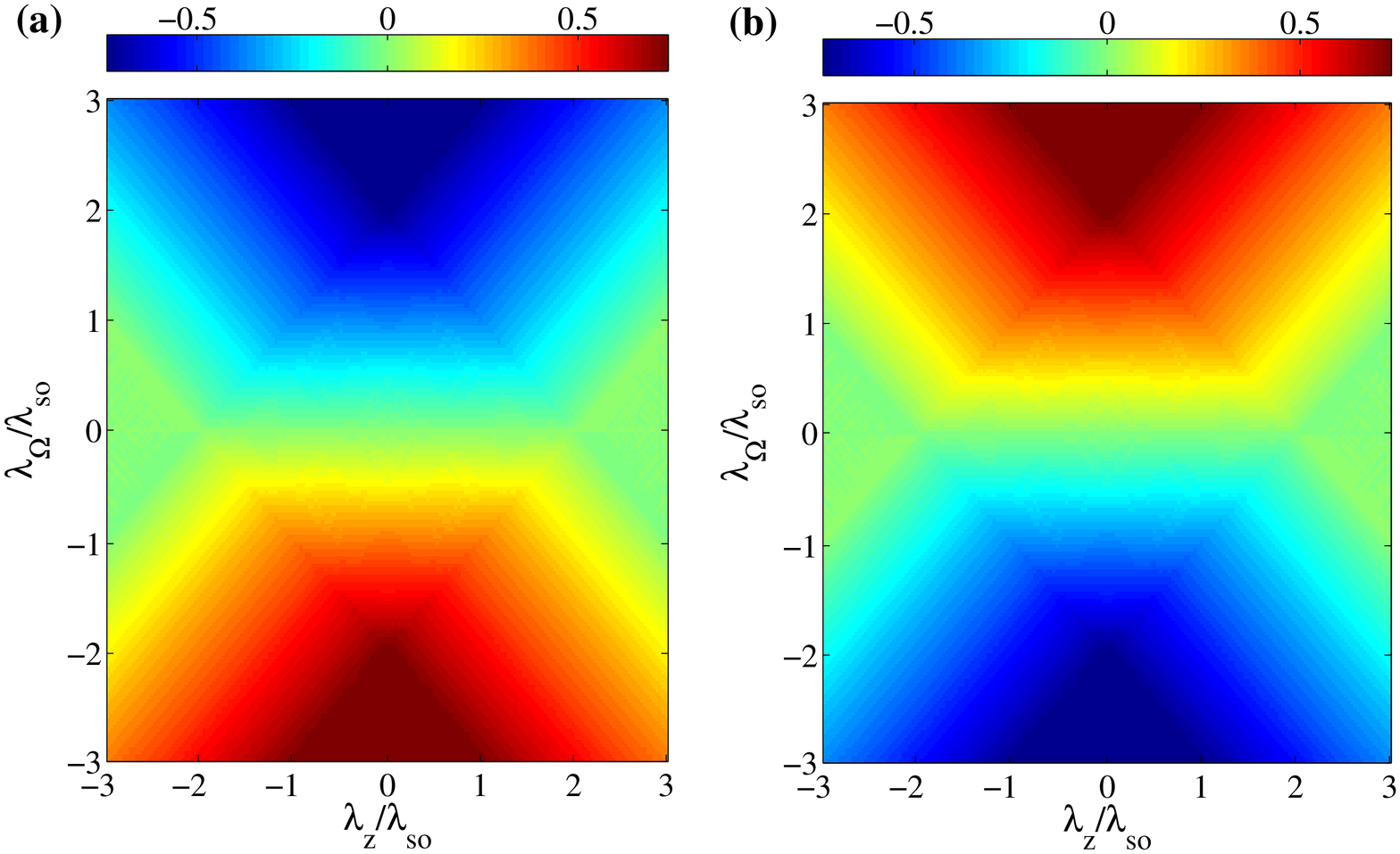}
\caption{Spin thermopower of an irradiated LBDM-based junction as
functions of $\lambda_{\Omega}/\lambda_{so}$ and
$\lambda_{z}/\lambda_{so}$ for (a) $\mu=-0.5\lambda_{so}$ and (b)
$\mu=+0.5\lambda_{so}$. The other parameters are
$T=0.4\lambda_{so}/k_{B}$ and $L=130a$.}\label{Fig07}
\end{center}
\end{figure}
\begin{figure}
\begin{center}
\includegraphics[width=15cm,angle=0]{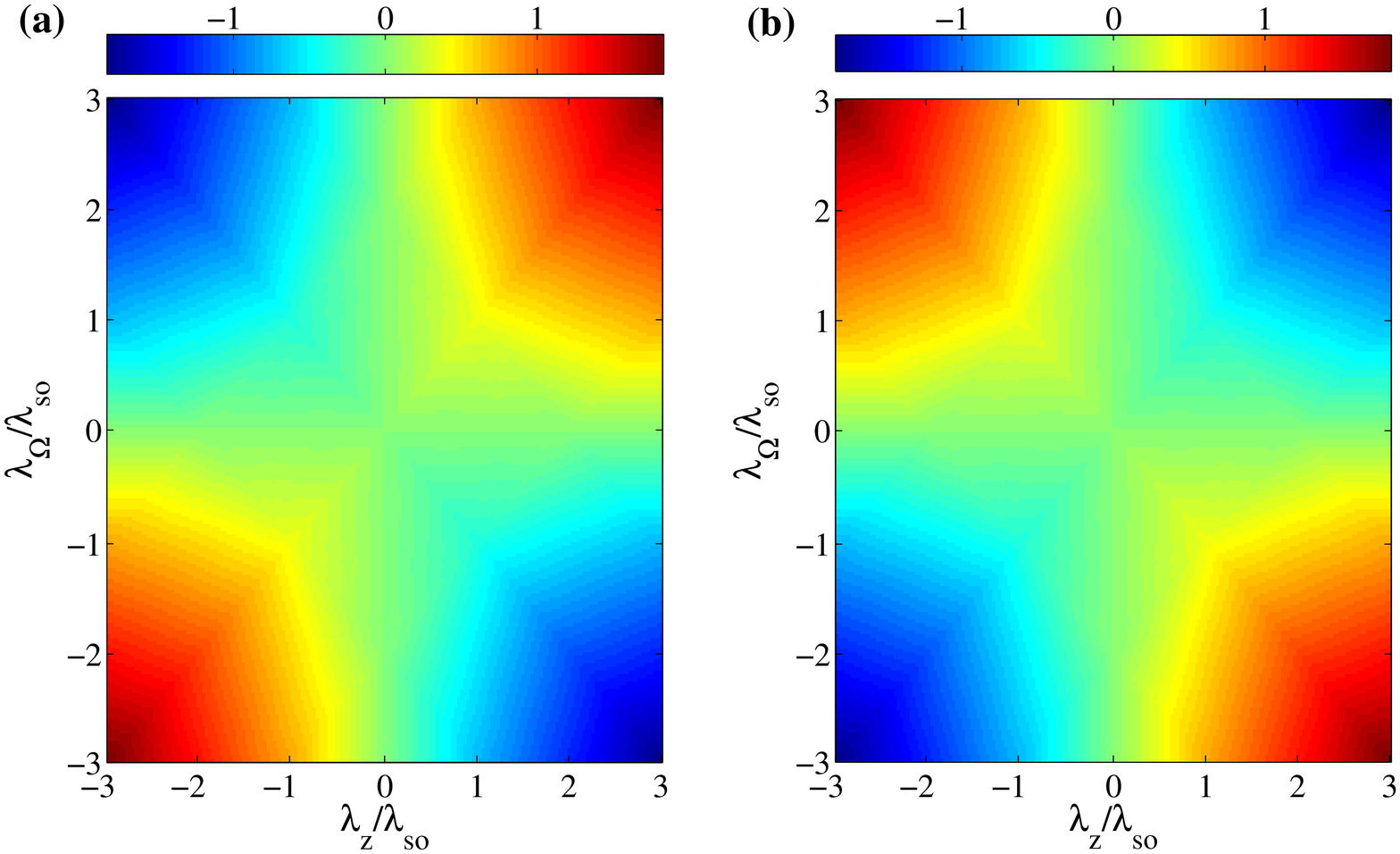}
\caption{Valley thermopower of an irradiated LBDM-based junction
as functions of $\lambda_{\Omega}/\lambda_{so}$ and
$\lambda_{z}/\lambda_{so}$ for (a) $\mu=-0.5\lambda_{so}$ and (b)
$\mu=+0.5\lambda_{so}$. The other parameters are
$T=0.4\lambda_{so}/k_{B}$ and $L=130a$.}\label{Fig08}
\end{center}
\end{figure}
\end{document}